\documentclass[twocolumn,secnumarabic,amssymb,amsmath,nofootinbib,
tightenlines,nobibnotes,aps,prl,showpacs,english]{revtex4}
\usepackage{graphicx,epsfig}
\begin{document}
\def\gsi{$^{a}$}
\def\hei{$^{b}$}
\def\mpi{$^{c}$}
\def\wei{$^{d}$}
\def\sun{$^{e}$}
\def\umn{$^{f}$}
\def\dub{$^{g}$}
\def\bnl{$^{h}$}

\title{Semi-Hard Scattering Unraveled from Collective Dynamics 
by Two-Pion Azimuthal Correlations in~158~A~GeV/$c$ Pb~+~Au Collisions}

\author{G.~Agakichiev\gsi, 
H.~Appelsh\"auser\hei,
R.~Baur\hei,
J.~Bielcikova\hei$^,$\mpi,
P.~Braun-Munzinger\gsi, 
A.~Cherlin\wei, 
A.~Drees\sun, 
S.\,I.~Esumi\hei, 
K.~Filimonov\hei,
Z.~Fraenkel\wei,
Ch.~Fuchs\mpi, 
P.~Gl\"assel\hei, 
G.~Hering\gsi,
P.~Huovinen\umn,
B.~Lenkeit\hei, 
A.~Mar\'{\i}n\gsi, 
F.~Messer\sun,\mpi, 
M.~Messer\hei,
J.~Milosevic\hei,
D.~Mi\'skowiec\gsi,
O.~Nix\mpi,
Yu.~Panebrattsev\dub, 
V.~Petr\'a\v{c}ek\hei,
A.~Pfeiffer\hei, 
J.~Rak\mpi, 
I.~Ravinovich\wei, 
S.~Razin\dub,
P.~Rehak\bnl, 
H.~Sako\gsi,
N.~Saveljic\dub,
W.~Schmitz\hei, 
S.~Shimansky\dub, E.~Socol\wei,
H.\,J.~Specht\hei, J.~Stachel\hei, 
H.~Tilsner\hei, 
I.~Tserruya\wei, 
C.~Voigt\hei,
S.~Voloshin\hei,
C.~Weber\hei,
J.\,P.~Wessels\hei, 
J.\,P.~Wurm\mpi, 
V.~Yurevich\dub~~(CERES Collaboration).\\
}

\address{
\gsi \mbox{Gesellschaft f\"ur Schwerionenforschung (GSI),
64291 Darmstadt, Germany}\\
\hei \mbox{Physikalisches Institut, Heidelberg University,
69120 Heidelberg, Germany}\\
\mpi \mbox{Max-Planck Institut f\"ur Kernphysik, 69229 Heidelberg,
Germany} \\
\wei \mbox{Weizmann Institute, Rehovot 76100, Israel}\\
\sun \mbox{Department of Physics and Astronomy, State University
of New York, Stony Brook, New York 11974, U.S.A.}\\
\umn \mbox{School of Physics and Astronomy, University of Minnesota,
Minneapolis, Minnesota 55455, U.S.A.}\\
\dub \mbox{Laboratory for High Energy (JINR), 141980 Dubna, Russia}\\
\bnl \mbox{Brookhaven National Laboratory, Upton, New York 11793-5000,
U.S.A.}\\
}
        
\begin{abstract} 
  Elliptic flow and two-particle azimuthal correlations of charged
  hadrons and high-$p_T$ pions ($p_T>$~1~GeV/$c$) have been measured
  close to mid-rapidity in 158A~GeV/$c$ Pb+Au collisions by the CERES
  experiment. Elliptic flow ($v_2$) rises linearly with $p_T$ to a value 
  of about 10$\%$ at 2~GeV/$c$. Beyond $p_T\approx$~1.5~GeV/$c$, 
  the slope decreases considerably, possibly indicating 
  a saturation of $v_2$ at high $p_T$.  Two-pion azimuthal anisotropies 
  for $p_T>$~1.2~GeV/$c$ exceed the
  elliptic flow values by about 60$\%$ in mid-central collisions. These
  non-flow contributions are attributed to near-side and back-to-back
  jet-like correlations, the latter exhibiting centrality dependent
  broadening.
\end{abstract}

\maketitle 
Late stages of ultra-relativistic heavy-ion collisions are
characterized by strong collective transverse expansion.  An important
signature of collective dynamics in non-central collisions is {\it
elliptic flow} manifesting itself in an azimuthal anisotropy of
particle yields with respect to the reaction
plane~\cite{Ollitrault:1992bk,Voloshin:2002wa}.  It is driven by
anisotropic pressure gradients built up during the early stage of the
collision in the geometrically anisotropic overlap
zone~\cite{Sorge:1998mk}. The degree of equilibration achieved during
the subsequent evolution might in principle be gauged by comparison to
hydrodynamic models which indicate an upper limit of elliptic flow.
Its magnitude depends on initial conditions and the equation of state
(EoS), but also on system lifetime~\cite{Teaney:2000cw}.  It had been
argued~\cite{Kolb:2000fh} that hydrodynamics was unable to describe
elliptic flow data at SPS energies while it accurately describes RHIC
data~\cite{AA}.

Two-particle azimuthal correlations are sensitive to collective flow
but might reveal, particularly at large transverse momentum ($p_T$),
also relics of primary scattering in the semi-hard sector which are
masked in inclusive $p_T$~spectra at SPS energies by the Cronin
effect~\cite{Wang:1998hs}. It is the purpose of this Letter to
demonstrate that collective flow and semihard scattering can be
disentangled by measurements of azimuthal anisotropies. We present
data (i) of elliptic flow for charged particles and for identified
pions up to $p_T\approx 3~$GeV/$c$, and (ii) of two-pion correlations
at $p_T\geq$~1.2~GeV/$c$. We pursue a statistical
analysis~\cite{Slivova:2002wj} which conjectures that the observed
anisotropies of non-flow nature are due to dijet-like correlations. An
interpretation in terms of resonance decays is unlikely in view of the
high invariant mass ($\approx$~2.5~GeV/$c^2$) required.

CERES\,\cite{Agakishiev:1998wu-Lenkeit:1999xu} has acceptance close to
mid-rapidity (2.1~$\leq$~$\eta$~$\leq$~2.65) with full azimuthal
coverage. Charged particles ($h^\pm$) are tracked by a doublet of
silicon-drift detectors (SDD) before, and a multiwire proportional
chamber behind the magnetic field used for momentum determination.
Identification of pions with $p>$~4.5~GeV/$c$ is performed by two
ring-imaging Cherenkov detectors~(RICH).  Pion momenta are determined
from the ring radii.  We analyzed 43~million 158~AGeV/$c$ Pb+Au
collisions taken in 1996 at the most central (26$\pm1.5)\%$ of the
geometric cross section, $\sigma_{geo}$. We divide the triggered
events into six contiguous centrality bins labeled by the percentage
of $\sigma_{geo}$ of their upper edges, 26$\%$, 21$\%$, 17$\%$,
13$\%$, 9$\%$, and 5$\%$, respectively. The centrality is determined
by charged particle multiplicity $\langle N_{ch}\rangle$ in the range
2~$\le\eta\le$~3 covered by the SDDs.  Numbers of participants,
$N_{part}$, and binary collisions, $N_{coll}$, were calculated from a
nuclear overlap model~\cite{Eskola:1989yh} neglecting fluctuations,
resulting in a total inelastic cross section of 6.94~barn.

Conventional elliptic flow analysis is based on the azimuthal particle
distribution with respect to the orientation $\Psi$ of the
reconstructed event plane (EP),
\begin{equation}
\frac{dN}{d(\phi-\Psi)}=
A[~1+2v_2^{\prime}\cos(2(\phi-\Psi))~].
\end{equation}
Measurements of the EP and of the particle anisotropy
are obtained from different subsets of the same
data sample: the $\phi$ acceptance is divided into 100 slices and every
fourth slice is combined into a subsample. We avoid autocorrelations
by using non-adjacent $\phi$-slices for $v_2^\prime$ and EP
measurements.  Non-uniformities in $\Psi$ are removed by standard
procedures~\cite{Poskanzer:1998yz}.
\begin{figure}[t!]
\includegraphics{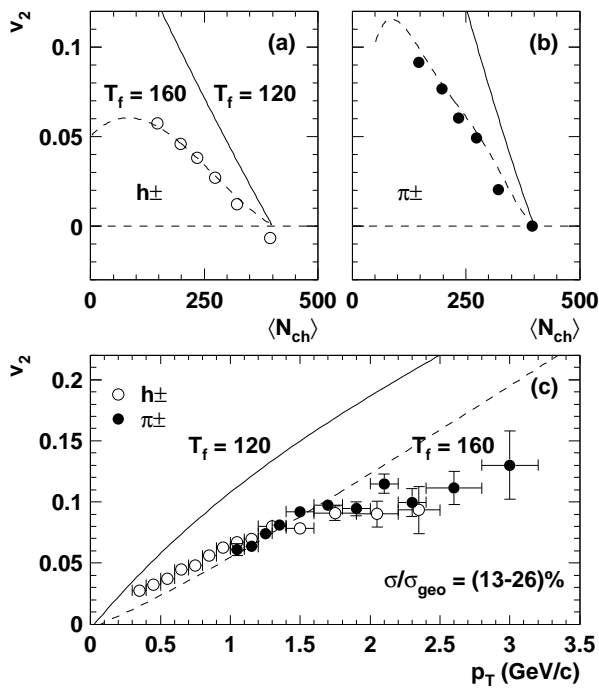}
\caption{ Centrality dependence of $v_2$ for (a) h$^\pm$,
0.5~$<p_T<$~2.5~GeV/$c$, and (b) $\pi^\pm$,
$p_T>$~1.2~GeV/$c$. Statistical errors are within symbols. (c)
$p_T$ dependence of $v_2$, corrected for BEC effects. Here, the
centrality corresponds to three left-most bins in (a),(b)
combined.  Hydrodynamical calculations with
phase transition at $T_c$= 165~MeV are shown for
kinetic freeze-out temperatures of 120~MeV and 160~MeV.}
\label{v2nchpt}
\end{figure}
Depending on centrality, the r.m.s. of EP resolution is 35-40~degrees.
The Fourier coefficient $v_2$ is obtained after correcting for the EP
dispersion, $v_2={\cal{K}}~v_2^{\prime}$.  The correction factor for
sample $i$, ${\cal K}_i=\langle
\mathrm{cos}(2(\Psi^i-\Psi_R))\rangle^{-1}$, involves the unknown
angle $\Psi_R$ of the true reaction plane but can be expressed by the
measured sample differences $\Psi^i-\Psi^j$. This spread in $\cal{K}$
factors dominates the systematic uncertainties in $v_2$. Resulting
absolute systematic errors vary with centrality between 0.5$\%$ and
1.5$\%$.

The flow results are presented in Fig.~\ref{v2nchpt}. With centrality,
$v_2$ decreases almost linearly as expected from hydrodynamics. The
larger $v_2$ values for $\pi^\pm$ reflect their larger $\langle
p_T\rangle\approx$~1.45~GeV/$c$ compared to 0.70~GeV/$c$ for
$h^\pm$. The $p_T$ dependence of $v_2$ measured for the first time at
the SPS up to 3~GeV/$c$ is shown in Fig.~\ref{v2nchpt}(c).  The data
are averaged over the three most peripheral centrality bins and
correspond to (13-26)$\%$ of $\sigma_{geo}$ ($\langle
N_{ch}\rangle=$~190).  It is seen that $v_2$ rises about linearly with
$p_T$ to a value of about 10$\%$ at 2~GeV/$c$. Beyond
$p_T\approx$~1.5~GeV/$c$, the slope decreases considerably, possibly
indicating a saturation of $v_2$ at high $p_T$ as observed at
RHIC~\cite{AA}.
\begin{figure}[b!]
\includegraphics{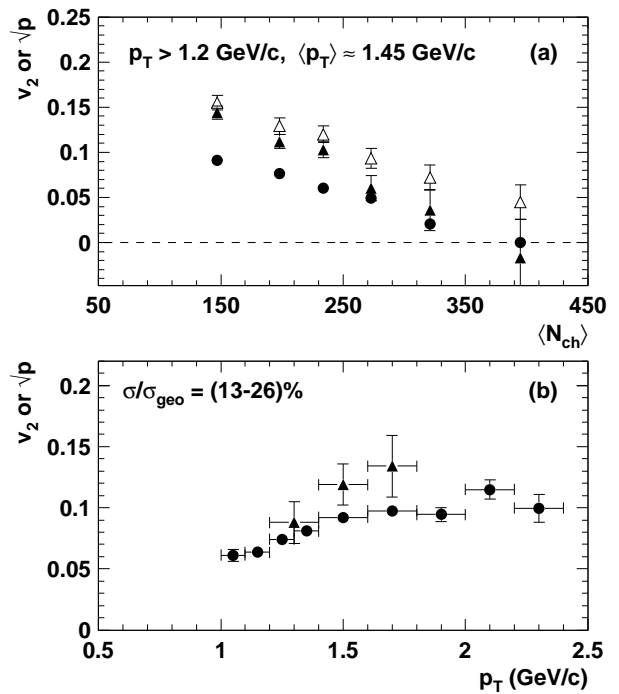}
\caption{Anisotropies $v_2$ from EP method for pions (circles) and 
  $\sqrt{p}$ from two-pion correlations (triangles). (a) Centrality 
  dependence for full azimuth (open triangles) and a range restricted 
  to $|\Delta\phi|\geq$~0.6~rad (closed triangles). 
  (b) $p_T$ dependence of $v_2$ for the top (13-26)$\%$ of 
  $\sigma_{geo}$, $\langle N_{ch}\rangle=$~190. }
\label{v2-rp-cor}
\end{figure} 

The data in Fig.~\ref{v2nchpt}(c) have been corrected for the effects
of Bose-Einstein correlations (BEC)~\cite{Dinh:1999mn} with input
from~\cite{Adamova:2002wi}. The corrections vary between -15$\%$ of
$v_2$ at $p_T=$~0.25~GeV/$c$ and +10$\%$ at $p_T>$~1~GeV/$c$.  Since
the procedure becomes questionable for central collisions, the data in
Figs.~\ref{v2nchpt}(a,b) were left uncorrected.  We may compare to
recent results of NA49~\cite{NA49} at (12.5-33.5)$\%$. After
correcting for different centrality, these are still about 15$\%$
larger in the range 1.0$\leq p_T\leq$1.5~GeV/$c$.

Results of hydrodynamical calculations~\cite{Kolb:2000fh} using an EoS
with a first order phase transition to quark gluon plasma at
$T=$~165~MeV and freeze-out at $T_f=$~120~MeV are considerably above
the data.  If the hydrodynamical evolution is terminated already at
$T_f=$~160~MeV~\cite{florkowski}, good agreement with the elliptic
flow data is reached. It should be noted, though, that in this case
the $p_T$ spectra of protons are
too steep (not shown). Conversely, the $T_f$= 120~MeV calculation
reproduces the spectra. Possible explanations include incomplete
thermalization~\cite{Teaney01} and viscous effects~\cite{Teaney02}.

We turn to the measurement of two-pion angular correlations at
$p_T\geq 1.2$~GeV/$c$ which are written in terms of the relative
azimuthal angle $\Delta\phi$ as
\begin{eqnarray}
\frac{dN}{d\Delta\phi}=
B~[1+ 2p\cos(2\Delta\phi)].
\end{eqnarray}
From the data we determine the second Fourier coefficient $p$ which
for pure flow is equal to $v_2^2$~\cite{Poskanzer:1998yz}.  The
two-pion yield is corrected for single-track reconstruction efficiency
which is determined by embedding simulated tracks into real events.
This correction varies between 16$\%$ and 9$\%$, depending on
$N_{ch}$.  At small opening angles, overlapping rings in the RICHes
cause a drop in pair reconstruction efficiency. To be less sensitive
to a Monte-Carlo (MC) correction, pairs with track separation
$\Delta\theta\leq$~20~mrad in polar angle were discarded.  This cut
reduces the pair efficiency loss by a factor of four while keeping
still 60$\%$ of statistics.  Corrected opening-angle distributions
reveal strong a strong anisotropy with maxima at $\Delta\phi\approx 0$
and $\pi$ (see below).  The procedure is supported by the fact that
anisotropies remain essentially unchanged after correction whether or
not the $\Delta\theta$ cut is applied.
\begin{figure}[b!]
\includegraphics{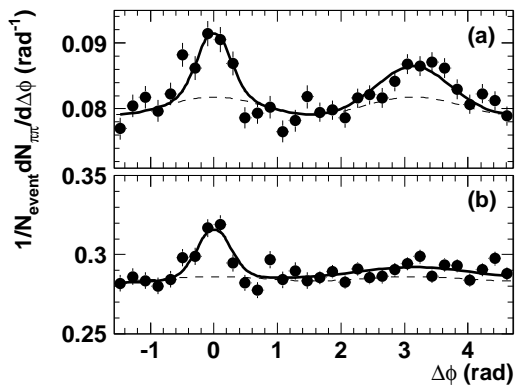}
\caption{
  Two-pion opening angle distributions for $p_T>~$1.2~GeV/$c$ (a) for
  the first centrality bin, (21-26)$\%$ and (b) for the fourth,
  (9-13)$\%$.  A cut $\Delta\theta\geq$~20~mrad and corrections for
  close-pair efficiency losses are applied.  Full line shows Gaussian
  fits to semi-hard components on top of flow-modulated background
  (dashed line).}
\label{dip-and-gauss}
\end{figure}

In Fig.~\ref{v2-rp-cor}(a) is shown that the $\sqrt{p}$ values from
the two-particle correlation are systematically larger than the $v_2$
coefficients. It is unlikely that the $v_2$ values are significantly
reduced by a possible bias on EP reconstruction by the high-$p_T$
particles in view of their very small abundance ($\approx$10$^{-3}$ of
$h^{\pm}$).  We have analyzed the full azimuth applying the
$\Delta\theta$ cut and MC correction for pair efficiency, and
alternatively a range $|\Delta\phi|\geq$~0.6~rad without these
remedies. For the first centrality bin ((21-26)$\%$, $\langle
N_{ch}\rangle=$~147), the value of $\sqrt{p}$ exceeds $v_2$ by 70$\%$
for full range in azimuth; accounting for BEC effects, this excess is
reduced to about 60$\%$.  The anisotropy in the restricted
$\Delta\phi$ range is similar in magnitude in the first bin as for the
full range, but it decreases more strongly with centrality and
approaches zero for central collisions ($<$5$\%$, $\langle
N_{ch}\rangle=~$395).
The gap between anisotropies from two-pion correlation and
conventional flow widens with increasing $p_T$ as can be seen
from Fig.~\ref{v2-rp-cor}(b). However, the statistical accuracy is
significantly degraded by invoking a two-dimensional window in $p_T$.

\begin{figure}[t!]
\includegraphics{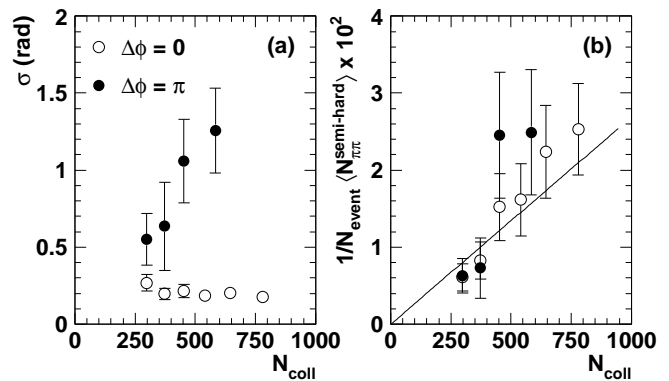}
\caption{ Centrality dependence of the Gaussian widths of the
correlation peaks at $\Delta\phi=~0,~\pi$ (a) and of the areas under
the Gaussian peaks, both from fits (b). The most central points for
$\Delta\phi=~\pi$ combine the fourth and fifth centrality bin,
(5-13)$\%$.  The loss in pair acceptance due to the cut
$\Delta\theta\geq$~20~mrad has not been corrected for. }
\label{broadening}
\end{figure}

The observed excess is attributed to direct pion-pion correlations,
presumably of semi-hard origin, in addition to collective flow. The
$\Delta\phi$ distributions shown in Fig.~\ref{dip-and-gauss}(a,b) are
well described by two Gaussians at $\Delta\phi$~=~0, $\pi$ on top of
elliptic flow.  Fit parameters are the Gaussian amplitudes and widths
and background $B$, while $v_2$ is fixed independently by the EP
method ($v_2^2({\rm EP})$ replacing $p$ in Eq.(2)). The results in
Fig.~\ref{broadening}(a) show that the close-angle peak stays narrow
at $\sigma_{0}=$~(0.23$\pm$0.03)~rad, consistent with
fragmentation~\cite{Arnison:1983}. The back-to-back peak broadens with
centrality up to $\sigma_{\pi}=$~(1.26$\pm$0.28)~rad at (5-13)$\%$,
from where on it cannot even be discerned from background. Within the
statistical errors, the yield contained in both peaks grows linearly
with $N_{coll}$ which supports the suggested interpretation of
semi-hard scattering. So, the back-to-back component escapes detection
in central collisions due to broadening but does not appear to be
suppressed in yield.

Partonic rescatterings in medium cause an imbalance in $p_T$
perpendicular to the initial hard scattering
plane~\cite{Appel-Blaizot:1986,Kopeliovich:1995} leading to $p_T$
broadening that is reflected in the width of the back-to-back
peak~\cite{Corcoran:1991vq}.  In contrast, the close-angle peak is not
affected since both pions originate from fragmentation of the same
parton, and propagating as color singlets experience only little
rescattering thereafter.  In a small-angle approximation, the $p_T$
broadening is
\mbox{$(\Delta p_T^2)^{1/2}\approx\langle p_T\rangle(\sigma^2_{\pi}-
\sigma^2_{0})^{1/2}$}.  For  (5-13)$\%$ centrality we
obtain $(\Delta p_T^2)^{1/2}$~=~(1.8$\pm$0.4)~GeV/$c$. A more accurate
treatment~\cite{Rak:2003} yields (2.8$\pm$0.6)~GeV/$c$. Both estimates
are similar to values measured in pA
collisions~\cite{Corcoran:1991vq}.

\begin{figure}[t!]
\includegraphics{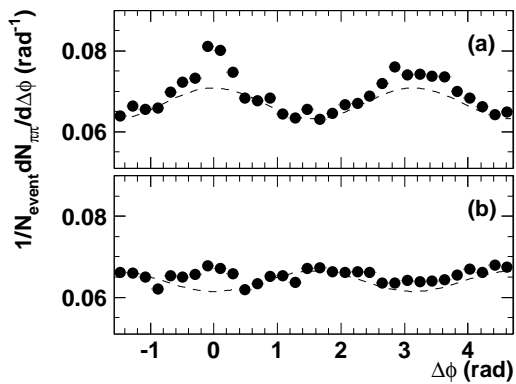}
\caption{In-plane (a) and out-of-plane (b) two-pion opening angle
distributions.  Dashed lines are calculated for pure
elliptic flow as measured by the EP method and corrected for BEC.
Data are for centrality 13-26$\%$, $p_T\geq$~1.2~GeV/$c$,
a cut on $\Delta\theta\geq$~20~mrad, and are efficiency corrected.}
\label{inout}
\end{figure}

Strong non-flow contributions to the two-particle opening-angle
anisotropy are confirmed by distributions in which one
of the two pions is detected in the EP ($\pm\pi/4$), or
perpendicular to it (again $\pm\pi/4$).  The anisotropy,
calculated~\cite{Slivova03} for pure elliptic flow as
\begin{equation}
p_{in/out}= v_2\cdot\left \langle{\cos(2(\phi-
\Psi))}\right\rangle_{in/out},
\end{equation}
is shown by dashed lines in Fig.~\ref{inout}, using the measured
$v_2=$~8.5$\%$. The angular brackets indicate an average over the
respective in-plane and out-of-plane sectors, weighted with the
anisotropic yields of Eq.(1), and folded with the measured EP
resolution.

{\it In-plane} (Fig.~\ref{inout}(a)), elliptic flow and semi-hard pairs
both peak at $\Delta\phi$= 0, $\pi$ where we observe an {\it excess}
over flow.  {\it Out-of-plane} (Fig.~\ref{inout}(b)), the flow pattern
is shifted by $\pi/2$, and the semi-hard correlation (always peaking
at $\Delta\phi$=~0 and $\pi$) fills in the minima of flow. Would the
jet-like correlations actually be misidentified elliptic flow, then an
increased harmonic amplitude in the out-of-plane correlation would
show more {\it negative} swings at 0 and $\pi$ than the dashed line in
Fig.~\ref{inout}(b), which is not the case.

{\it Summary and discussion.}$-$ Differential $v_2(p_T)$ depends on
centrality and $p_T$ as expected from hydrodynamical calculations, but
we have not been able to reproduce both the magnitude of
our $v_2$ data and the $p_T$ spectra. Above 1.5~GeV/$c$ the slope of
measured $v_2$ flattens while the calculation continues to rise.

The observation of semi-hard two-particle azimuthal correlations
embedded in collective flow is novel for SPS energies. The broadening
of the back-to-back correlation with increasing centrality suggests
that we observe in-medium partonic scattering which affects both parts
of a dijet independently. Although the back-to-back correlation
broadens in central collisions, there is no sign of its suppression.
The absence of broadening of the close-angle correlation supports the
view that these pions originate from fragmentation of the same parton.
Our results thus exhibit similar features, but also important
differences, to recent findings at RHIC~\cite{Adler:2002ctq}.

We thank P.M.~Dinh and J.-Y.~Ollitrault for help, A.~Accardi, B.~Kopeliovich and E.~Shuryak for
enlightening discussions.  The work was supported by the German BMBF,
the U.S. DoE, the German Israeli and the Minerva Foundations, and the
Nella and Leon Benoziyo Center for High Energy Physics Research.

\end{document}